\documentclass[amsmath, aps]{revtex4}
\usepackage{graphicx}
\usepackage{times}
\usepackage{epsfig}
\usepackage{amssymb}
\usepackage{amsmath}
\begin{document}
\title{Spin-Statistic Selection Rules for Multiphoton Transitions: Application to Helium Atom}
\author{T. Zalialiutdinov$^{1}$, D. Solovyev$^{1}$, L. Labzowsky$^{1,2}$ and G. Plunien$^{3}$}

\affiliation{$^1$ Department of Physics, St.Petersburg State University, Ulianovskaya 1, Petrodvorets, St.Petersburg 198504, Russia
\\
$^2$  Petersburg Nuclear Physics Institute, 188300, Gatchina, St.Petersburg, Russia
\\
$^3$ Institut f\"{u}r Theoretische Physik, Technische Universit\"{a}t Dresden, Mommsenstrasse 13, D-10162, Dresden, Germany}

\begin{abstract}
A theoretical investigation of the three-photon transition rates $ 2^1P_1\rightarrow 2^1S_0\;,1^1S_0 $ and  $ 2^3P_2\rightarrow 2^1S_0\;,1^1S_0 $ for the helium atom is presented. Photon energy distributions and precise values of the nonrelativistic transition rates are obtained with employment of correlated wave functions of the Hylleraas type. The possible experiments for the tests of the Bose-Einstein statistics for multiphoton systems are discussed.
\end{abstract}

\maketitle
\section{Introduction}
This paper presents a theoretical investigation of the three-photon decay rates from $ 2^1P_1 $ and $ 2^3P_2 $ states in the helium atom. The primary motivation for these calculations is the recent series of spectroscopic tests of Bose-Einstein statistics for the system of two photons \cite{demille}-\cite{angom}. The experiments use a selection rule [2, 3] for atomic transitions that is closely related to the Landau-Yang theorem (LYT) \cite{landau, yang}. The selection rule states that two equivalents photons cannot participate in any process that would require them to be in a state with total angular momentum equal to one. An evident example in the high-energy physics is the prohibition of the two-photon decay for the neutral spin-one $Z^{0}$-boson. The same concerns the annihilation decay of orthopositronium (also spin-1 state). However both these decays are also forbidden by charge-parity conservation law. Positronium presents a real neutral system (it coincides with itself after charge conjugation), therefore it possesses a definite charge parity \cite{berlif}, connected with the total spin value $S$: parapositronium ($S=0$) is charge-positive and orthopositronium ($ S=1 $) is charge negative. Since the charge parity of a system of $ N_{\gamma} $ photons equals $ (-1)^{N_{\gamma}} $ \cite{akhiezer}, parapositronium can not decay into an odd number of (not necessarily equivalent) photons and orthopositronium can not decay into an even number  of photons. The $ Z^{0} $-boson as a charge-parity-negative particle can not decay into an even number of photons. Recently LYT was employed for the proof that the heavy particle discovered at LHC \cite{higgs} has spin $S=0$, i.e. presents Higgs boson. A similar situation exists for atomic transitions. The early calculations of the two-photon decay of the singlet $ 2^1S_0\equiv(1s2s)^1S_0 $ and triplet $ 2^3S_1\equiv(1s2s)^3S_1 $ excited states of He and He-like ions to the ground $ 1^1S_0\equiv(1s)^{2\;1}S_0 $ state revealed the crucial difference in the photon frequency distributions in both cases \cite{bely}, \cite{drakevictor}. The decay probability for the triplet case turns to zero when the frequencies of the emitted photons are equal. Later these conclusions were confirmed within the fully relativistic calculations (see, for example \cite{derevianko}). A neutral atom unlike positronium is not a real neutral system and does not possess a definite charge parity. Therefore the differencies in atomic transitions are exclusively due to the Bose-Einstein statistics.

In \cite{zlsg} the Spin-Statistics Selections Rules (SSSRs) for multiphoton transitions with equal photons in atomic systems were established which present an extension of LYT to the 3- and 4-photon systems. The SSSR-1 sounds like: two equivalent photons can not participate in any atomic transition that would require them to have total (common) odd angular momentum. The SSSR-2 is: three equivalent dipole photons can not participate in any atomic transition that would require  the to have total even angular momentum. The SSSR-3 claims that four equivalent dipole photons can not have total odd angular momentum. 

For the experimental tests of SSSRs it is possible to use laser source. Unlike original LYT formulation (this was mentioned both by Landau \cite{landau} and Yang \cite{yang}) the derivation \cite{zlsg} is valid also for the collinear i.e. for the laser photons. As particular examples the two-, three- and four-photon transitions in He-like U were considered in \citep{zlsg}. This choice was explained, from the one side by the relative simplicity of the calculations in two-electron ions with the full neglect of the interelectron interaction. This neglect does not change the SSSRs but makes the numerical results inaccurate. The inaccuracy drops down with increase of the nuclear charge and is minimal (about 1\%) for U. From the other side the spectrum in two-electron atoms is reach enough to provide all interesting multiphoton transitions. However such atomic system as He-like uranium are not suitable for possible experimental tests of SSSRs. For $ Z=92 $ the energy difference between states $ 2^1P_1 $, $ 2^3P_2 $ and ground state $ 1^1S_0 $ is of hundreds KeV and prevents the use of optical lasers. An advantage of the use of the laser source is that all the photons will have the same frequency. If we divide the energy interval $ \omega_a $ between appropriate atomic states by an integer number $N_{\gamma}$ and adjust the laser frequency $\omega_l$ to this value, $\omega_l=\omega_a/N_{\gamma}$, the number of photons $N_{\gamma}$ in the absorption process will be fixed. The value of the total angular momentum $J$ for $N_\gamma$-photon system can be fixed by choosing the appropriate values $J_{e_i}$ and $J_{e_f}$ for the initial (lower) and final (upper) atomic levels in the transition process. For example, if we choose $J_{e_i}=0$, $J_{e_f}=2$ and $N_{\gamma}=3$ the SSSR-2 can be tested for $N_{\gamma}=3, J=2$  which prohibits three dipole photons with the same parity and equal frequency to be in a quantum state with total angular momentum $ J=2 $. What we can not  fix is the multipolarity of photon, i.e. a total angular momentum $j$ of every separate photon. A laser light in the beam can be decomposed in all possible multipolarities. This means, for example that together with E1E1E1 transition, all transitions with the same total parity constructed with the higher multipoles, i.e. E1M1E2, E1E1M2 etc. will be always absorbed. However the processes with the photons of higher multipolarities are usually strongly suppressed in atoms, therefore the E1E1E1 transition will be dominant. Measuring the absorption rate at the $\omega_l=\omega_a/N_{\gamma}$ frequency one can establish the validity or non-validity of the particular SSSR: the atomic vapour should be transparent for the laser light at the frequency $\omega_l=\omega_a/N_{\gamma}$. Note also that unlike the spontaneous emission which is very weak for multiphoton transitions, the multiphoton absorption depends on the laser intensity and can be well observed in the experiments \cite{mokler}.

For the test of SSSR-2 one can choose, in particular the transition $ 2^1S_0\rightarrow 2^3P_2 $ with transition energy $ 0.35 $ eV. This transition should obey the SSSR-2, i.e. three-photon absorption of the laser photons with equal frequencies should be absent. For comparison it is convenient to employ $ 2^1S_0-2^1P_1 $ transition with the transition energy $ 0.60 $ eV, which should not exhibit such properties since in this case three equal photons should have the total angular momentum $ J=1 $. It is important that in both cases the  initial $ 2^1S_0 $ state is metastable what enables to perform laser absorption experiments.

The most advantageous for the experimental tests of SSSRs is neutral helium atom where transitions correspond to the optical or infrared region. For example for the test of SSSR-2 three-photon transitions between the $ 2^1S_0 $ state and excited $ 2^1P_1 $ and $ 2^3P_2 $ states can be used. Unlike $ U^{90+} $ ion the calculations for neutral He atom can be performed with high accuracy. The application of SSSRs to neutral He is the main purpose of the present paper. Accurate variational calculations of the three-photon transitions are given, based on the employment of the variational wave functions. Unlike the case of $ U^{90+} $ these calculations can be done with nonrelativistic wave functions (overall accuracy of about 0.01\%). A powerful method for obtaining the variational wave functions of Hylleraas type developed in \cite{kor_monk}-\cite{korobov1} was employed for these calculations.

\section{Theory}
We have calculated $ 2^3P_2\rightarrow 1^1S_0 $, $ 2^3P_2\rightarrow 2^1S_0 $,$ 2^1P_1\rightarrow 1^1S_0 $ and $ 2^1P_1\rightarrow 2^1S_0 $ three-photon transition rates. Transitions from $ 2^3P_2 $ level exhibit the behaviour governed by SSSR-2 and transitions from $ 2^3P_1 $ level were calculated for comparison. 

The three-photon decay of $ 2^1P_1 $ state is qualitatively different from $ 2^3P_2 $ three-photon decay since the SSSR-2 plays significant role in $ 2^3P_2\rightarrow 1^1S_0+3\gamma(E1) $ transition. This difference affects both the angular and the energy distributions of three photons. In \cite{zlsg} it was shown that in He-like uranium $ 2^3P_2\rightarrow 1^1S_0+3\gamma(E1) $ transition rate turns to zero when the frequencies of three photons are equal while $ 2^1P_1\rightarrow 1^1S_0+3\gamma(E1) $ transition rate has the maximum in the same situation. This is also true for the neutral helium because the angular part remains the same. 
 
Fully relativistic form of the differential transition rate in conjunction with the integration over photon directions $ \vec{\nu}=\vec{k}/|\vec{k}| $ and summation over the photon polarizations $ \vec{e} $ of all the photons with the account for all permutations of photons is  \cite{zlsg}
\begin{eqnarray}
\label{1}
\frac{dW_{i\rightarrow f}(\omega_1,\omega_2)}{d\omega_1d\omega_2} = \frac{\omega_3\;\omega_2\;\omega_1}{(2\pi)^5}\sum_{\lambda_3\lambda_2\lambda_1}\sum_{j_{\gamma_3}j_{\gamma_2}j_{\gamma_1}}\sum_{m_{\gamma_3}m_{\gamma_2}m_{\gamma_1}}
\\
\nonumber
\left|\sum\limits_{n'n}\frac{\left(Q^{(\lambda_3)}_{j_{\gamma_3}m_{\gamma_3}\omega_3}\right)_{fn'}\left(Q^{(\lambda_2)}_{j_{\gamma_2}m_{\gamma_2}\omega_2}\right)_{n'n}\left(Q^{(\lambda_1)}_{j_{\gamma_1}m_{\gamma_1}\omega_1}\right)_{ni}}{(E_{n'} - E_f - \omega_3) (E_n - E_f - \omega_3 - \omega_2)}+
\sum\limits_{n'n}\frac{\left(Q^{(\lambda_3)}_{j_{\gamma_3}m_{\gamma_3}\omega_3}\right)_{fn'}\left(Q^{(\lambda_1)}_{j_{\gamma_1}m_{\gamma_1}\omega_1}\right)_{n'n}\left(Q^{(\lambda_2)}_{j_{\gamma_2}m_{\gamma_2}\omega_2}\right)_{ni}}{(E_{n'} - E_f - \omega_3) (E_n - E_f - \omega_3 - \omega_1)}+
\right.
\\
\nonumber
\sum\limits_{n'n}\frac{\left(Q^{(\lambda_1)}_{j_{\gamma_1}m_{\gamma_1}\omega_1}\right)_{fn'}\left(Q^{(\lambda_3)}_{j_{\gamma_3}m_{\gamma_3}\omega_3}\right)_{n'n}\left(Q^{(\lambda_2)}_{j_{\gamma_2}m_{\gamma_2}\omega_2}\right)_{ni}}{(E_{n'} - E_f - \omega_1) (E_n - E_f - \omega_1 - \omega_3)}+
\sum\limits_{n'n}\frac{\left(Q^{(\lambda_1)}_{j_{\gamma_1}m_{\gamma_1}\omega_1}\right)_{fn'}\left(Q^{(\lambda_2)}_{j_{\gamma_2}m_{\gamma_2}\omega_2}\right)_{n'n}\left(Q^{(\lambda_3)}_{j_{\gamma_3}m_{\gamma_3}\omega_3}\right)_{ni}}{(E_{n'} - E_f - \omega_1) (E_n - E_f - \omega_1 - \omega_2)}+
\\
\nonumber
\sum\limits_{n'n}\frac{\left(Q^{(\lambda_2)}_{j_{\gamma_2}m_{\gamma_2}\omega_2}\right)_{fn'}\left(Q^{(\lambda_3)}_{j_{\gamma_3}m_{\gamma_3}\omega_3}\right)_{n'n}\left(Q^{(\lambda_1)}_{j_{\gamma_1}m_{\gamma_1}\omega_1}\right)_{ni}}{(E_{n'} - E_f - \omega_2) (E_n - E_f - \omega_2 - \omega_3)}+
\left.\sum\limits_{n'n}\frac{\left(Q^{(\lambda_2)}_{j_{\gamma_2}m_{\gamma_2}\omega_2}\right)_{fn'}\left(Q^{(\lambda_1)}_{j_{\gamma_1}m_{\gamma_1}\omega_1}\right)_{n'n}\left(Q^{(\lambda_3)}_{j_{\gamma_3}m_{\gamma_3}\omega_3}\right)_{ni}}{(E_{n'} - E_f - \omega_2) (E_n - E_f - \omega_2 - \omega_1)}\right|^2\;,
\end{eqnarray} 
where $ \omega_i\;(i=1,2,3) $ are the photon frequencies, $ j_{\gamma} $, $ m_{\gamma} $ are the photon angular momenta and their projections, $ \lambda_{i}=0,\;1,\;-1 $, the indices $ i,\;n,\;f $ represent the initial, intermediate and final atomic states respectively. Summation over $ n,\;n' $ in Eq. (\ref{1}) extends over entire atomic spectrum. The following notations are used \cite{shonin} 
\begin{eqnarray}
\label{2}
Q^{(0)}_{j_{\gamma}m_{\gamma}\omega}=\vec{\alpha} \vec{A}^{(0)}_{{j_\gamma}m_{\gamma}}(\vec{r})\;,
\end{eqnarray}
\begin{eqnarray}
\label{3}
Q^{(1)}_{j_{\gamma}m_{\gamma}\omega}=\vec{\alpha} \vec{A}^{(1)}_{{j_\gamma}m_{\gamma}}(\vec{r})+G_{\gamma}A^{(-1)}_{{\gamma}m_{\gamma}}(\vec{r})\;,
\end{eqnarray}
where $ \vec{A}^{(0)}_{{j_\gamma}m_{\gamma}}(\vec{r}) $, $ \vec{A}^{(1)}_{{j_\gamma}m_{\gamma}}(\vec{r}) $ are magnetic and electric vector potentials, $ \vec{A}^{(-1)}_{{j_\gamma}m_{\gamma}}(\vec{r}) $ is the scalar potential
\begin{eqnarray}
\label{4}
\vec{A}^{(0)}_{{j_\gamma}m_{\gamma}}(\vec{r})=i^{j_{\gamma}}g_{j_\gamma}(\omega r)\vec{Y}^{*}_{j_\gamma j_\gamma m_\gamma}(\vec{n})\;,
\end{eqnarray}
\begin{eqnarray}
\label{5}
\vec{A}^{(1)}_{{j_\gamma}m_{\gamma}}(\vec{r})=i^{j_{\gamma}+1}\left\lbrace \sqrt{\frac{j_\gamma}{2j_\gamma+1}}g_{j_\gamma+1}(\omega r)\vec{Y}^{*}_{j_\gamma j_\gamma+1 m_\gamma}(\vec{n})-\sqrt{\frac{j_\gamma+1}{2j_\gamma+1}}g_{j_\gamma-1}(\omega r)\vec{Y}^{*}_{j_\gamma j_\gamma-1 m_\gamma}(\vec{n})\right.\\\nonumber+\left.G_{j_\gamma}\left(\sqrt{\frac{j_\gamma+1}{2j_\gamma+1}}g_{j_\gamma+1}(\omega r)\vec{Y}^{*}_{j_\gamma j_\gamma+1 m_\gamma}(\vec{n})+\sqrt{\frac{j_\gamma}{2j_\gamma+1}}g_{j_\gamma-1}(\omega r)\vec{Y}^{*}_{j_\gamma j_\gamma-1 m_\gamma}(\vec{n})\right)\right\rbrace\;,
\end{eqnarray}
\begin{eqnarray}
\label{6}
A^{(-1)}_{{j_\gamma}m_{\gamma}}(\vec{r})=i^{j_{\gamma}}g_{j_\gamma}(\omega r)Y^{*}_{j_\gamma m_\gamma}(\vec{n})\;.
\end{eqnarray}
\begin{eqnarray}
\label{7}
g_{j_\gamma}=4\pi j_{j_\gamma}(\omega r)
\end{eqnarray}
Here $ G_{j_{\gamma}} $ is the gauge parameter defining gauge for the electromagnetic potentials, $  j_{j_{\gamma}}(kr) $ is a spherical Bessel function, $ \vec{n}=\frac{\vec{r}}{\left|r\right|} $. 

Actually for our purposes (calculations for He atom) the fully nonrelativistic calculation of transition energies as well as fully nonrelativistic expression for the photon emission operators are quite sufficient. We have only to take into account the spin-orbit interaction for describing the intercombination transitions.

In the nonrelativistic limit ($ kr\ll 1 $) with  $G_{j_{\gamma}}=\sqrt{\frac{j_{\gamma}+1}{j_{\gamma}}}$ and $ j_{\gamma}=1 $ for the dipole photons Eq. (\ref{3}) reduces to the form
\begin{eqnarray}
\label{8}
Q^{(1)}_{1m_{\gamma}\omega}=i\frac{4\sqrt{2}}{3}\pi \omega r Y^{*}_{1m_\gamma}(\vec{n})\;,
\end{eqnarray}
Using definition for the spherical components of vector $ \vec{r} $
\begin{eqnarray}
\label{9}
r^{m}_1=\sqrt{\frac{4\pi}{3}}\left|\vec{r}\;\right|Y^{*}_{1m}(\vec{n})
\end{eqnarray}
Eq. (\ref{8}) can be presented in the form
\begin{eqnarray}
\label{10}
Q^{(1)}_{1m_{\gamma}\omega}=i\;\sqrt{\frac{8\pi}{3}} \omega\;r^{m}_1\;.
\end{eqnarray}
Substitution of Eq. (\ref{10}) to the expression for the differential three-photon transition rate Eq. (\ref{1}) yields 
\begin{eqnarray}
\label{11}
\frac{dW_{i\rightarrow f}(\omega_1,\omega_2)}{d\omega_1d\omega_2} = \frac{16}{27\pi^2}\left(\omega_3\;\omega_2\;\omega_1\right)^3\sum_{m_{\gamma_3}m_{\gamma_2}m_{\gamma_1}}
\\
\nonumber
\left|\sum\limits_{n'n}\frac{\left(r^{m_{\gamma_3}}\right)_{fn'}\left(r^{m_{\gamma_2}}\right)_{n'n}\left(r^{m_{\gamma_1}}\right)_{ni}}{(E_{n'} - E_f - \omega_3) (E_n - E_f - \omega_3 - \omega_2)}+
\sum\limits_{n'n}\frac{\left(r^{m_{\gamma_3}}\right)_{fn'}\left(r^{m_{\gamma_1}}\right)_{n'n}\left(r^{m_{\gamma_2}}\right)_{ni}}{(E_{n'} - E_f - \omega_3) (E_n - E_f - \omega_3 - \omega_1)}+
\right.
\\
\nonumber
\sum\limits_{n'n}\frac{\left(r^{m_{\gamma_1}}\right)_{fn'}\left(r^{m_{\gamma_3}}\right)_{n'n}\left(r^{m_{\gamma_2}}\right)_{ni}}{(E_{n'} - E_f - \omega_1) (E_n - E_f - \omega_1 - \omega_3)}+
\sum\limits_{n'n}\frac{\left(r^{m_{\gamma_1}}\right)_{fn'}\left(r^{m_{\gamma_2}}\right)_{n'n}\left(r^{m_{\gamma_3}}\right)_{ni}}{(E_{n'} - E_f - \omega_1) (E_n - E_f - \omega_1 - \omega_2)}+
\\
\nonumber
\sum\limits_{n'n}\frac{\left(r^{m_{\gamma_2}}\right)_{fn'}\left(r^{m_{\gamma_3}}\right)_{n'n}\left(r^{m_{\gamma_1}}\right)_{ni}}{(E_{n'} - E_f - \omega_2) (E_n - E_f - \omega_2 - \omega_3)}+
\left.\sum\limits_{n'n}\frac{\left(r^{m_{\gamma_2}}\right)_{fn'}\left(r^{m_{\gamma_1}}\right)_{n'n}\left(r^{m_{\gamma_3}}\right)_{ni}}{(E_{n'} - E_f - \omega_2) (E_n - E_f - \omega_2 - \omega_1)}\right|^2\;.
\end{eqnarray} 
In the $ N_{e} $-electron atom $ r^{m_{\gamma}}=\sum\limits_{i=1}^{N_{e}}r^{m_{\gamma}}_i $, where $ r^{m_{\gamma}}_i  $ are the spherical components of radius-vector for $ i $-th electron. Then the total transition rate can be defined as 
\begin{eqnarray}
\label{12}
W_{i\rightarrow f}=\frac{1}{3!}\frac{1}{2j_{ei}+1}\sum\limits_{m_{ei},m_{ef}}\;\iint \frac{dW_{i\rightarrow f}(\omega_1,\omega_2)}{d\omega_1d\omega_2}d\omega_1d\omega_2\;.
\end{eqnarray}
Summation over all projections appearing in the expression (\ref{11}) can be performed numerically for each value of corresponding angular momenta. 

To perform the numerical calculations of three-photon transitions in two-electron atomic systems we use the variational wave functions of Hylleraas type \cite{korobov1}. The wave function for a state with a total electron orbital angular momentum $ L_{e} $, its projection $ M_{e} $ and total spatial parity $ \pi_{e}=(-1)^{L_{e}} $ is expanded as follows
\begin{eqnarray}
\label{13}
\Psi_{L_{e}M_{e}}\left(\vec{r}_1\vec{r}_2\right)=\sum\limits_{l_{e_1}+l_{e_2} = L_{e}}\left[Y^{L_{e}M_{e}}_{l_{e_1}l_{e_2}}\left(\vec{n}_1,\vec{n}_2\right)G_{l_{e_1}l_{e_2}}^{L_{e}\pi_{e}}\left(r_1,r_2\right)\pm(1\leftrightarrow 2)\right]\;,
\end{eqnarray} 
where $ G_{l_{e_1}l_{e_2}}^{L_{e}\pi_{e}} $ is the radial part, corresponding to a certain bipolar harmonics $ Y^{L_{e}M_{e}}_{l_{e_1}l_{e_2}} $ \cite{varsh}. These function are obtained by the variational method developed in \cite{kor_monk}-\cite{korobov1}. The method consists of expansion of $ G_{l_{e_1}l_{e_2}}^{L_{e}\pi_{e}} $ in the exponential basis set with a complex parameters $ \alpha_{i} $ , $ \beta_{i} $ and $ \gamma_{i} $  generated in a quasirandom manner \cite{kor_monk} with the size of basis defined by integer number $ N $
\begin{eqnarray}
\label{14}
G_{l_{e_1}l_{e_2}}^{L_{e}\pi_{e}}\left(r_1,r_2\right)=\sum\limits_{i=1}^{N}\left\lbrace U_{i}\;\mbox{Re}\left[\mbox{exp}\left(-\alpha_{i}r_1-\beta_{i}r_2-\gamma_{i}r_{12}\right)\right]+W_{i}\;\mbox{Im}\left[\mbox{exp}\left(-\alpha_{i}r_1-\beta_{i}r_2-\gamma_{i}r_{12}\right)\right]\right\rbrace
\;,
\end{eqnarray}
where $ r_{12}=\left|\vec{r_1}-\vec{r_2}\right| $, $ U_{i} $ and $ W_{i} $ are linear parameters. 

Since we are interested in the transitions between states with a certain values of total momentum $ \vec{J}_e=\vec{L}_e+\vec{S_e} $, the $ L_eS_eJ_eM_e $ coupling scheme needs to be used, where $ \vec{S_e} $ is the total spin value \cite{berlif,varsh}. Then 
\begin{eqnarray}
\label{16}
\left(r^{m_{\gamma}}\right)_{n'L_{e}'S_{e}'J_{e}'M_{e}'nL_{e}S_{e}J_{e}M_{e}}=\delta_{S_{e}'S_{e}}\left(-1\right)^{J_{e}'-M_{e}'}
\begin{pmatrix}
 J_{e}' & 1 & J_{e}\\
-M_{e}' & m_{\gamma} & M_{e}
\end{pmatrix}\\\nonumber
\times\sqrt{(2J_{e}'+1)(2J_{e}+1)}(-1)^{L_{e}'+S_{e}'+J_{e}+1}
\begin{Bmatrix}
L_{e}' & J_{e}' & S_{e}'\\
J_{e}  & L_{e}  & 1
\end{Bmatrix}
\langle n'L_{e}'\left|\left|r\right|\right| nL_{e}\rangle
\end{eqnarray}
Evaluation of reduced matrix element in Eq. (\ref{16}) on the basis of Hylleraas-type wave functions is described in \cite{drake1}.

As an illustration of SSSR-2 for three equal photons we consider $2^{3}P_2\rightarrow 1^1S_0+3\gamma(E1)$ transition in the helium atom and demonstrate that the value $J=2$ of the total angular momentum for the three equivalent dipole photons is prohibited. In principle, this decay can proceed via several channels. First, $2^3P_2\rightarrow n^3S_1+\gamma(E1)\rightarrow n'^3P_1\left[n'^1P_1\right]+2\gamma(E1)\rightarrow 1^1S_0+3\gamma(E1)$; this channel is prohibited since the transition $n^3S_1\rightarrow 1^1S_0+2\gamma(E1)$ is prohibited by SSSR-1. Second, $2^3P_2\rightarrow n^3D_1+\gamma(E1)\rightarrow n'^3P_1\left[n'^1P_1\right]+2\gamma(E1)\rightarrow 1^1S_0 +3\gamma(E1)$; this channel is also prohibited by SSSR-1 since the transition $ n^3D_1\rightarrow 1^1S_0 +2\gamma(E1)$ is prohibited by this rule. The contribution of the third channel  $2^3P_2 \rightarrow n^3D_2\left[n^1D_2\right]+\gamma(E1)\rightarrow n'^3P_1\left[n'^1P_1\right]+2\gamma(E1)\rightarrow 1^1S_0+3\gamma(E1)$ does not turn to zero so evidently. The states admixed by the spin-orbit interaction are placed in the square brackets, $n,\, n'$ are sequential numbers for the states with the same symmetry in two-electron atoms.

The $2^{3}P_2\rightarrow 1^1S_0+3\gamma(E1)$ transition unlike the $2^{1}P_1\rightarrow 1^1S_0+3\gamma(E1)$ transition proceeds only through spin-orbit mixing of the intermediate states. Following \cite{mathis, drakevictor} we write the true $ P $ and $ D $ wave functions in the form
\begin{eqnarray}
\label{17}
\left|n'\;^3P_1\right.\rangle_{\mathrm{true}}=\left|n'\;^3P_1\right.\rangle + \sum\limits_{n''=2}^{\infty}\epsilon^{(P)}_{n'n''}\left|n''\;^1P_1\right.\rangle\;,
\end{eqnarray}
\begin{eqnarray}
\label{18}
\left|n'\;^1P_1\right.\rangle_{\mathrm{true}}=\left|n'\;^1P_1\right.\rangle - \sum\limits_{n''=2}^{\infty}\epsilon^{(P)}_{n''n'}\left|n''\;^3P_1\right.\rangle\;,
\end{eqnarray}
\begin{eqnarray}
\label{19}
\left|n'\;^3D_2\right.\rangle_{\mathrm{true}}=\left|n'\;^3D_2\right.\rangle + \sum\limits_{n''=3}^{\infty}\epsilon^{(D)}_{n'n''}\left|n''\;^1D_2\right.\rangle\;,
\end{eqnarray}
\begin{eqnarray}
\label{20}
\left|n'\;^1D_2\right.\rangle_{\mathrm{true}}=\left|n'\;^1D_2\right.\rangle - \sum\limits_{n''=3}^{\infty}\epsilon^{(D)}_{n''n'}\left|n''\;^3D_2\right.\rangle\;,
\end{eqnarray}
where
\begin{eqnarray}
\label{21}
\epsilon^{(P)}_{n'n''}=\frac{\langle n'\;^3P_1\left| H_{3}\right| n''\;^1P_1\rangle}{E(n'\;^3P)-E(n''\;^1P)}\;,
\end{eqnarray}
\begin{eqnarray}
\label{22}
\epsilon^{(D)}_{n'n''}=\frac{\langle n'\;^3D_2\left| H_{3}\right| n''\;^1D_2\rangle}{E(n'\;^3D)-E(n''\;^1D)}
\end{eqnarray}
and $ H_{3} $ is the spin-orbit interaction operator \cite{bethe}. In the absence of external electric and  magnetic fields, the lowest-order spin-dependent relativistic corrections $ H_{3} $ consist of the spin-orbit term (in atomic units)
\begin{eqnarray}
\label{23}
H_{\mathrm{so}}=\frac{Z}{2c^2}\sum\limits_{i}\left[\frac{\vec{r}_i\times\vec{p}_i}{r_i^3}\right]\cdot\widehat{s}_i
\end{eqnarray}
and the spin-other-orbit term
\begin{eqnarray}
\label{24}
H_{\mathrm{soo}}=\frac{1}{2c^2}\sum\limits_{i\neq j}\left[\frac{\vec{r}_{ij}\times\vec{p}_i}{r_{ij}^3}\right]\cdot\left(\widehat{s}_i+2\;\widehat{s}_{j}\right)\;.
\end{eqnarray}
Matrix elements of Eqs. (\ref{23}) and (\ref{24}) can be reduced to \cite{drake3, alexander}
\begin{eqnarray}
\label{25}
\langle L_{e}S_{e}'J_{e}M_{e}\left|H_{\mathrm{so}}\right| L_{e}S_{e}J_{e}M_{e}\rangle = \frac{2}{c^2}(-1)^{L_{e}+S_{e}'+J_{e}}
\begin{Bmatrix}
J_{e}  & S_{e}' & L_{e}\\
1  & L_{e}  & S_{e}
\end{Bmatrix}
\left\langle L_{e}\left|\left|\frac{\vec{r}_1\times\vec{p}_1}{r_1^3}\right|\right| L_{e}\right\rangle \left\langle S_{e}'\left|\left|\widehat{s}_1\right|\right| S_{e}\right\rangle\;,
\end{eqnarray}

\begin{eqnarray}
\label{26}
\langle L_{e}S_{e}'J_{e}M_{e}\left|H_{\mathrm{soo}}\right| L_{e}S_{e}J_{e}M_{e}\rangle = \frac{1}{c^2}(-1)^{L_{e}+S_{e}'+J_{e}}
\begin{Bmatrix}
J_{e}  & S_{e}' & L_{e}\\
1  & L_{e}  & S_{e}
\end{Bmatrix}
\left\langle L_{e}\left|\left|\frac{\vec{r}_{12}\times\vec{p}_1}{r_{12}^3}\right|\right| L_{e}\right\rangle \left\langle S_{e}'\left|\left|\widehat{s}_1+2\widehat{s}_2\;\right|\right| S_{e}\right\rangle\;.
\end{eqnarray} 
Evaluation of reduced matrix elements in Eqs. (\ref{25})-(\ref{26}) on the Hylleraas type wave functions was presented in \cite{drakevictor}.
In this way for the $2^{3}P_2\rightarrow 1^1S_0+3\gamma(E1)$ transition probability we obtain
\begin{eqnarray}
\label{27}
\frac{dW_{i\rightarrow f}(\omega_1,\omega_2)}{d\omega_1d\omega_2} = \frac{16}{27\pi^2}\left(\omega_3\;\omega_2\;\omega_1\right)^3\sum_{m_{\gamma_3}m_{\gamma_2}m_{\gamma_1}}\left|\sum\limits_{n''\;n'\;n}\left\lbrace \left(U^{\;\gamma_1,\gamma_2,\gamma_3}_{\;n''n'n} + D^{\;\gamma_1,\gamma_2,\gamma_3}_{\;n''n'n}\right) +\mbox{5 permutations}\right\rbrace\right|^2\;,
\end{eqnarray}

\begin{eqnarray}
\label{28}
U^{\;\gamma_1,\gamma_2,\gamma_3}_{\;n''n'n}=\left(r^{m_{\gamma_3}}\right)_{1^1S_0\;n''}\epsilon^{\;ST}_{n''\;n'}\left(r^{m_{\gamma_2}}\right)_{n'\;n}\left(r^{m_{\gamma_1}}\right)_{n\;2^3P_2}\times
\\\nonumber\left\lbrace\frac{1}{(E^{\;S}_{n''} - E_f - \omega_3) (E^{\;T}_n - E_f - \omega_3 - \omega_2)}-\frac{1}{(E^{\;T}_{n'} - E_f - \omega_3) (E^{\;T}_n - E_f - \omega_3 - \omega_2)}\right\rbrace\;,
\end{eqnarray}

\begin{eqnarray}
\label{29}
D^{\;\gamma_1,\gamma_2,\gamma_3}_{\;n''n'n}=\left(r^{m_{\gamma_3}}\right)_{1^1S_0\;n''}\left(r^{m_{\gamma_2}}\right)_{n''\;n'}\epsilon^{\;ST}_{n'\;n}\left(r^{m_{\gamma_1}}\right)_{n\;2^3P_2}\times
\\\nonumber\left\lbrace \frac{1}{(E^{\;S}_{n''} - E_f - \omega_3) (E^{\;S}_{n'} - E_f - \omega_3 - \omega_2)}-\frac{1}{(E^{\;S}_{n''} - E_f - \omega_3) (E^{\;T}_n - E_f - \omega_3 - \omega_2)}\right\rbrace\;.
\end{eqnarray}
Here $ E^{S}_{n} $ and $ E^{T}_{n} $ are the energies of the singlet or triplet $ n $-states respectively. Permutations in Eq. (\ref{27}) are understood as permutations of the indices $1,2,3$. From Eq. (\ref{27}) it is clear that the probability of $ 2^3P_2 $ decay is suppressed by the smallness of the spin-orbit interaction. The same was observed also for the two-photon decay of $ 2^3S_1 $ state \cite{drakevictor} and, in principle, should hold for any intercombination transitions. 

For the $2^{3}P_2\rightarrow 1^1S_0+3\gamma(E1)$ transition in the channel $2^3P_2\rightarrow n^3S_1+\gamma(E1)\rightarrow n'^3P_1\left[n'^1P_1\right]+2\gamma(E1)\rightarrow 1^1S_0+3\gamma(E1)$ there is the resonance (the situation when the energy denominator turns to zero). The presence of the cascade-producing state $2^3S_1$ in the sum over all the intermediate states in the transition amplitude Eq. (\ref{27}) leads to the arrival of the high, but narrow "ridge" in the frequency distribution $\frac{dW(\omega_1,\omega_2)}{d\omega_1d\omega_2}$. Analogous situation arises for $2^{1}P_1\rightarrow 1^1S_0+3\gamma(E1)$ transition. However the existence of this "ridge" does not influence the validity of the SSSR-2, since "ridge" corresponds to the nonequal frequencies for the all 3-photon.

\section{Technical details and results}
Transition probabilities obtained from the expression Eq. (\ref{12}) were checked for the convergence with a different length of the basis set $ N $. Results of calculations with the $ N=30,\;50,\;100\;,150 $ for the $ 2^3P_2\rightarrow 2^1S_0+3\gamma(E1) $ and $ 2^1P_1\rightarrow 2^1S_0+3\gamma(E1) $ transitions are presented in Table I.   Corresponding plot for the $ 2^3P_2\rightarrow 2^1S_0+3\gamma(E1) $ transition is presented in Fig. 1. However it is more convenient to present 2-dimensional sectional cut of Fig. 1, with the fixed frequency of the third photon at the point $ \omega_3=\Delta/3 $, where $ \Delta=E(2^3P_2)-E(2^1S_0) $. This two-dimensional cut is presented in Fig. 2. Two-dimensional plots for the  $ 2^3P_2\rightarrow 1^1S_0+3\gamma(E1) $ and $ 2^1P_1\rightarrow 1^1S_0+3\gamma(E1) $ transition are presented in Fig. 3 and Fig. 4 respectively. The main difference between Fig. 3 and Fig. 4, arises at the point with coordinates $ \omega_1=\omega_2=\Delta/3 $ in the frequency distribution. For the transition $ 2^3P_2\rightarrow 1^1S_0+3\gamma(E1) $ differential transition rate becomes zero due to SSSR-2. These two transitions are not suitable for experiments due to the transition frequencies which is out of the optical range, but are calculated for additional demonstration of SSSR-2.

In order to integrate over photon frequency, the $ 32 $ points of Gauss-Legendre quadrature method was employed. The numerical values for the spin-orbit (Eq. (\ref{25})) and spin-other-orbit (Eq. (\ref{26})) matrix elements for some lowest $ P $ and $ D $ states are presented in Table II and are in a good agreement with the previous calculations \cite{alexander}, \cite{drake3}. Nonrelativistic variational energies of bound states obtained and used in this work are presented in Table III. They are in a very good agreement (13 significant digits) with previous calculations. We should stress that for our purposes it is sufficient to employ nonrelativistic energy levels: they provide an overall accuracy in our calculations at the level $ 0.01\% $.

In \cite{zlsg} we suggested a method for a possible experimental test of the SSSRs by means of laser light absorption and presented numerical examples with the highly charged He-like ions. The photon frequencies in this case are in the X-ray region which makes it difficult to perform the tests. In the present paper we suggest the helium atom as the most adequate system for the test of SSSRs with the optical lasers. The helium atom has long served as a testing ground for both theoretical and experimental studies and several reviews have chronicled the progress in both areas (see, for example, \cite{review}). The numerical calculations of three-photon $ 2^3P_2\rightarrow 2^1S_0+3\gamma(E1) $ and $ 2^1P_1\rightarrow 2^1S_0+3\gamma(E1) $ transitions in helium are listed in Table I. These transitions are most suitable for the laser testing of SSSR-2 due to the values of the corresponding transition frequencies (and hence to laser ones after dividing by $ N_{\gamma} $), belonging to the optical or infrared region.

Corresponding one-photon decay rates,  are also presented and are in good agreement with previous calculations \cite{drake5, alexander2} (see Table I). Our results coincide with the numbers given in \cite{alexander2} with 4 significant figures which means the accuracy of $ 0.01\% $. The same accuracy should have the 3-photon transition probabilities.

From the energy differences listed in Table I one can see that proposed transitions are suitable for the test of SSSR-2 since each emitted (absorbed) photon is in the optical (infrared) range. As it was mentioned in the introduction the corresponding one-photon decay is suppressed at the frequency $ \omega_l $. Note, that the energy difference between triplet and singlet $ P $ states in He atom are resolvable for the laser source.  

\begin{center}
Acknowledgments
\end{center}
The authors are indebted to V. I. Korobov for the permission to use the computer codes for the construction of the He variational wave functions and for valuable consultations. This work was supported by RFBR (grant No. 14-02-00188). T. Z., D. S. and L. L. acknowledge the support by St.-Petersburg State University with a research grant 11.38.227.2014. The work of T. Z. was also supported by grant of German-Russian Interdisciplinary Science Center (G-RISC) P-2015b-10 and non-profit foundation "Dynasty" (Moscow).

\begin{table}[hbtp]
\caption{Probabilities for the three-photon $ 2^3P_2\rightarrow 2^1S_0+3\gamma(E1) $, $ 2^1P_1\rightarrow 2^1S_0+3\gamma(E1) $ and one-photon $ 2^3P_2\rightarrow 2^1S_0+1\gamma(M2) $, $ 2^1P_1\rightarrow 2^1S_0+1\gamma(E1) $  transitions in $s^{-1}$. The number in parentheses indicates the power of ten. Transition energies $ \Delta E $ in eV are listed in two last columns.}
\begin{tabular}{ c  c  c | l | c | c | c}
\\
\hline
\hline
 $ N $ & $ W^{3\gamma(E1)}_{2^3P_2-2^1S_0} $ & $ W^{3\gamma(E1)}_{2^1P_1-2^1S_0} $ & $ W^{1\gamma(M2)}_{2^3P_2-2^1S_0}  $ & $ W^{1\gamma(E1)}_{2^1P_1-2^1S_0}  $ & $ \Delta E_{2^3P-2^1S} $ & $ \Delta E_{2^1P-2^1S} $
\qquad\\
\hline 
$ 30 $ & $ 3.727(-27) $ & $ 4.430(-15) $ & $  $ & $  $ & $  $ & $  $
\qquad\\
$ 50 $ & $ 3.763(-27) $ & $ 6.651(-15) $ & $  $ & $  $ & $  $ & $  $
\qquad\\
$ 100 $ & $ 3.763(-27) $ & $ 6.575(-15) $ & $  $ & $  $ & $ 0.35 $ & $ 0.60 $
\qquad\\
$ 150 $ & $ 3.764(-27) $ & $ 6.573(-15) $ & $  $ & $  $ & $  $ & $  $
\qquad\\
$ 500 $ & $ - $ & $ - $ & $ 1.1953(-8) $ & $ 1.9746(6)  $ & $  $ & $  $
\qquad\\
$ 	  $ & $   $ & $   $ & $ 1.1952(-8)\;[27],\;[28] $ & $  $ & $  $
\qquad\\
\hline 
\hline
\end{tabular}
\end{table}

\begin{table}[hbtp]
\caption{Values of singlet-triplet (ST) spin-orbit and spin-other-orbit matrix elements computed with $ N=500 $ (in a.u.)}
\begin{tabular}{ c  l  l }
\\
\hline
\hline
 $ \mbox{State} $ & \multicolumn{1}{c}{$\langle H_{\mathrm{so}}\rangle_{\mathrm{ST}}$}  & \multicolumn{1}{c}{$ \langle H_{\mathrm{soo}}\rangle_{\mathrm{ST}}  $ }
\qquad\\
\hline 
$ 2P $ & $ -0.000\;001\;9049679 $ & $ -0.000\;000\;688\;044 $
\qquad\\
$    $ & $ -0.000\;001\;907\;\mbox{\cite{alexander}} $ & $ -0.000\;000\;689\;\mbox{\cite{alexander}}  $
\qquad\\
$    $ & $ -0.000\;001\;904968\;\mbox{\cite{drake3}} $ & $ -0.000\;000\;688043\;\mbox{\cite{drake3}} $
\qquad\\
$ 3P $ & $ -0.000\;000\;558\;930 $ & $ -0.000\;000\;191\;302 $
\qquad\\
$    $ & $ -0.000\;000\;5604\;\mbox{\cite{alexander}} $ & $ -0.000\;000\;192\;6\;\mbox{\cite{alexander}} $ 
\qquad\\
$    $ & $ -0.000\;000\;558\;929\;\mbox{\cite{drake3}} $ & $ -0.000\;000\;191\;301\;\mbox{\cite{drake3}} $ 
\qquad\\
$ 4P $ & $ -0.000\;000\;234\;46 $ & $ -0.000\;000\;078\;773 $
\qquad\\
$    $ & $ -0.000\;000\;234\;6\;\mbox{\cite{alexander}} $ & $ -0.000\;000\;079\;3\;\mbox{\cite{alexander}} $ 
\qquad\\
$    $ & $ -0.000\;000\;234\;440\;\mbox{\cite{drake3}}    $ & $  -0.000\;000\;078\;772\;\mbox{\cite{drake3}} $ 
\qquad\\
$ 3D $ & $ -0.000\;000\;162\;89 $ & $ -0.000\;000\;080 $
\qquad\\
$    $ & $ -0.000\;000\;162\;70\;\mbox{\cite{alexander}} $ & $ -0.000\;000\;079\;96\;\mbox{\cite{alexander}} $ 
\qquad\\
$    $ & $ -0.000\;000\;162\;750 \;\mbox{\cite{drake3}} $ & $ -0.000\;000\;079\;992\;\mbox{\cite{drake3}} $ 
\qquad\\
\hline
\hline 
\end{tabular}
\end{table}

\begin{table}
\caption{Nonrelativistic energies for bound states of helium in a.u.}
\begin{tabular}{ c  l  l  l  l  l  l }
\\
\hline
\hline
$ \mbox{State}/N $ & \multicolumn{1}{c}{$ 30 $} & \multicolumn{1}{c}{$ 50 $} & \multicolumn{1}{c}{$ 100 $} & \multicolumn{1}{c}{$ 150 $} & \multicolumn{1}{c}{$ 500 $} & \multicolumn{1}{c}{$ \mbox{Drake}\;\mbox{\cite{drake_book}} $} 
\qquad\\
$ 1^1S $ & $ -2.903\;701 $ & $ -2.903\;722 $ & $ -2.903\;724\;357 $ & $ -2.903\;724\;371\;62 $ & $ -2.903\;724\;377\;034\;044 $ & $ -2.903\;724\;377\;034\;119\;5 $ 
\qquad\\
$ 2^3S $ & $ -2.175\;178 $ & $ -2.175\;225 $ & $ -2.175\;229\;365 $ & $ -2.175\;229\;377\;45 $ & $ -2.175\;229\;378\;236\;739\; $ & $ -2.175\;229\;378\;236\;791\;30 $ 
\qquad\\
$ 2^1S $ & $ -2.144\;864 $ & $ -2.145\;681 $ & $ -2.145\;973\;79 $ & $ -2.145\;973\;98 $ & $ -2.145\;974\;046\;042 $ & $ -2.145\;974\;046\;054\;419 $ 
\qquad\\
$ 2^3P $ & $ -2.132\;72 $ & $ -2.133\;14 $ & $ -2.133\;164 $ & $ -2.133\;164\;171 $ & $ -2.133\;164\;190\;756 $ & $ -2.133\;164\;190\;779\;273 $ 
\qquad\\
$ 2^1P $ & $ -2.123\;819 $ & $ -2.123\;839 $ & $ -2.123\;843\;017 $ & $ -2.123\;843\;067 $ & $ -2.123\;843\;086\;358 $ & $ -2.123\;843\;086\;498\;093 $ 
\qquad\\
$ 3^3D $ & $ -2.055\;521 $ & $ -2.055\;634 $ & $ -2.055\;636$ & $ -2.055\;636\;027 $ & $ -2.055\;636\;047 $ & $ -2.055\;636\;309\;453\;261 $ 
\qquad\\
$ 3^1D $ & $ -2.055\;553 $ & $ -2.055\;606 $ & $ -2.055\;618\;92 $ & $ -2.055\;618\;95 $ & $ -2.055\;618\;97 $ & $ -2.055\;620\;732\;852\;246 $ 
\qquad\\
\hline
\hline
\end{tabular}
\end{table}

\begin{figure}[hbtp]
\caption{(Color Online) 3-dimensional plot for frequencies distribution of the transition rate $ 2^3P_2\rightarrow 2^1S_0+3\gamma(E1) $ in Helium. On the vertical axis the transition rate $ \frac{dW}{d\omega_1d\omega_2}$ in $ s^{-1} $ is plotted; on the horizontal axes the photon frequencies are plotted in units $ \omega_1/\Delta $, $ \omega_2/\Delta  $ where  $ \Delta $ denotes the energy difference $ E(2^3P)-E(2^1S) $.  The lowest (zero) point is the point with coordinates $ \omega_1/\Delta=\omega_2/\Delta=1/3 $ at the bottom of the "pit" in the frequency distribution for the transition rate. This "pit" arises due to SSSR-2.}
\centering
\includegraphics[scale=0.57]{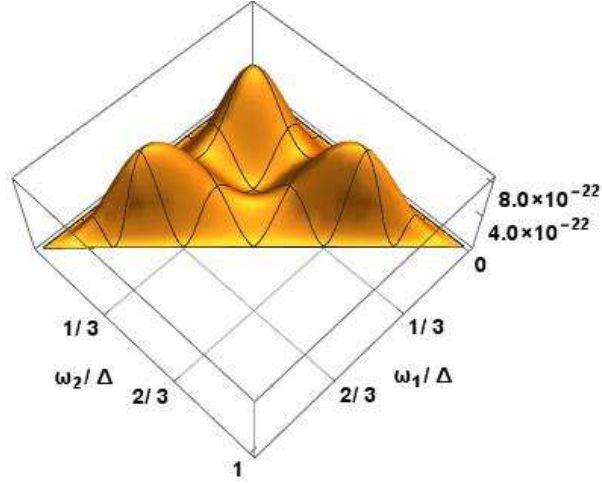}
\end{figure}
\begin{figure}[hbtp]
\caption{(Color Online) 2-dimensional sectional cut for frequencies distribution of the transition rate $ 2^3P_2\rightarrow 2^1S_0+3\gamma(E1) $ in Helium. On the vertical axis the transition rate $ \frac{dW}{d\omega_1}$ in $ s^{-1} $ is plotted; on the horizontal axis the photon frequency is plotted in units $ \omega_1/\Delta $  where  $ \Delta=E(2^3P)-E(2^1S) $. Second frequency $ \omega_2$ is fixed at the point $\omega_2/\Delta=1/3$. The lowest (zero) point is the point with coordinates $ \omega_1/\Delta=\omega_2/\Delta=1/3 $ at the bottom of the "pit" in the frequency distribution for the transition rate. This "pit" arises due to SSSR-2.}
\centering
\includegraphics[scale=0.6]{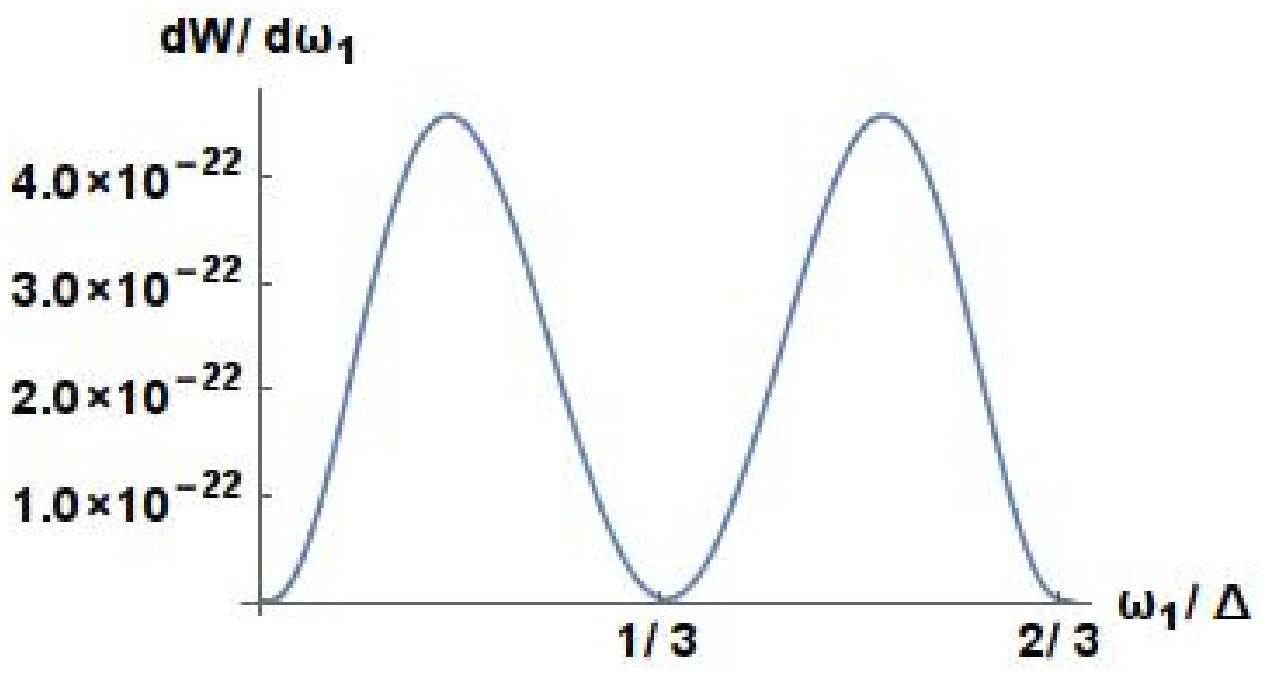}
\end{figure}

\begin{figure}[hbtp]
\caption{(Color Online) 2-dimensional sectional cut for frequencies distribution of the transition rate $ 2^3P_2\rightarrow 1^1S_0+3\gamma(E1) $ in Helium.  On the vertical axis the transition rate $ \frac{dW}{d\omega_1}$ in $ s^{-1} $ is plotted; on the horizontal axis the photon frequency is plotted in units $ \omega_1/\Delta $  where  $ \Delta=E(2^3P)-E(1^1S) $. Second frequency $ \omega_2$ is fixed at the point $\omega_2/\Delta=1/3$. The lowest (zero) point is the point with coordinates $ \omega_1/\Delta=\omega_2/\Delta=1/3 $ at the bottom of the "pit" in the frequency distribution for the transition rate. This "pit" arises due to SSSR-2.}
\centering
\includegraphics[scale=0.53]{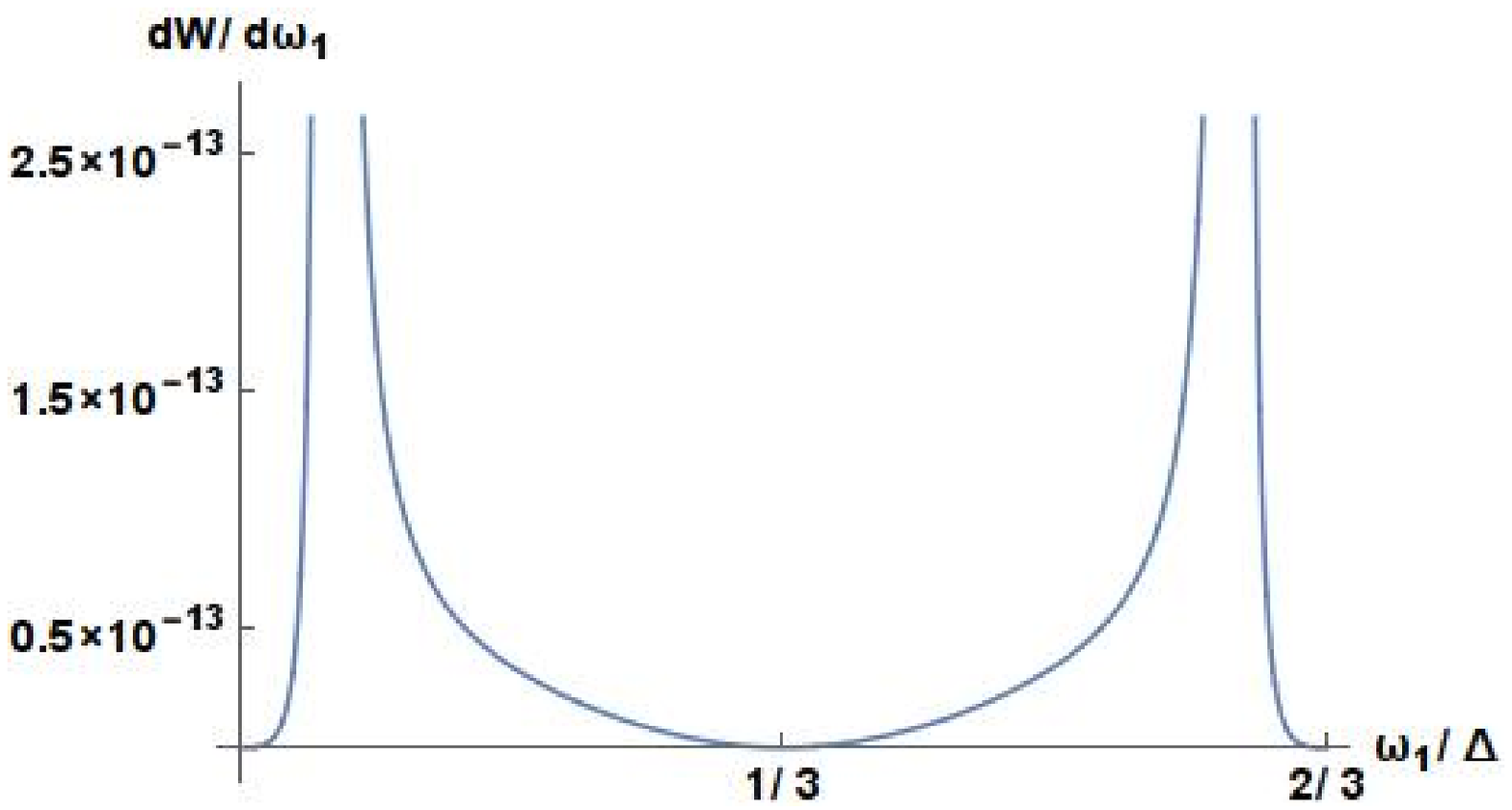}
\end{figure}
\begin{figure}[hbtp]
\caption{(Color Online) 2-dimensional sectional cut for frequencies distribution of the transition rate $ 2^1P_1\rightarrow 1^1S_0+3\gamma(E1) $ in Helium.  On the vertical axis the transition rate $ \frac{dW}{d\omega_1}$ in $ s^{-1} $ is plotted; on the horizontal axis the photon frequency is plotted in units $ \omega_1/\Delta $  where  $ \Delta=E(2^1P)-E(1^1S) $. Second frequency $ \omega_2$ is fixed at the point $\omega_2/\Delta=1/3$. }
\centering
\includegraphics[scale=0.57]{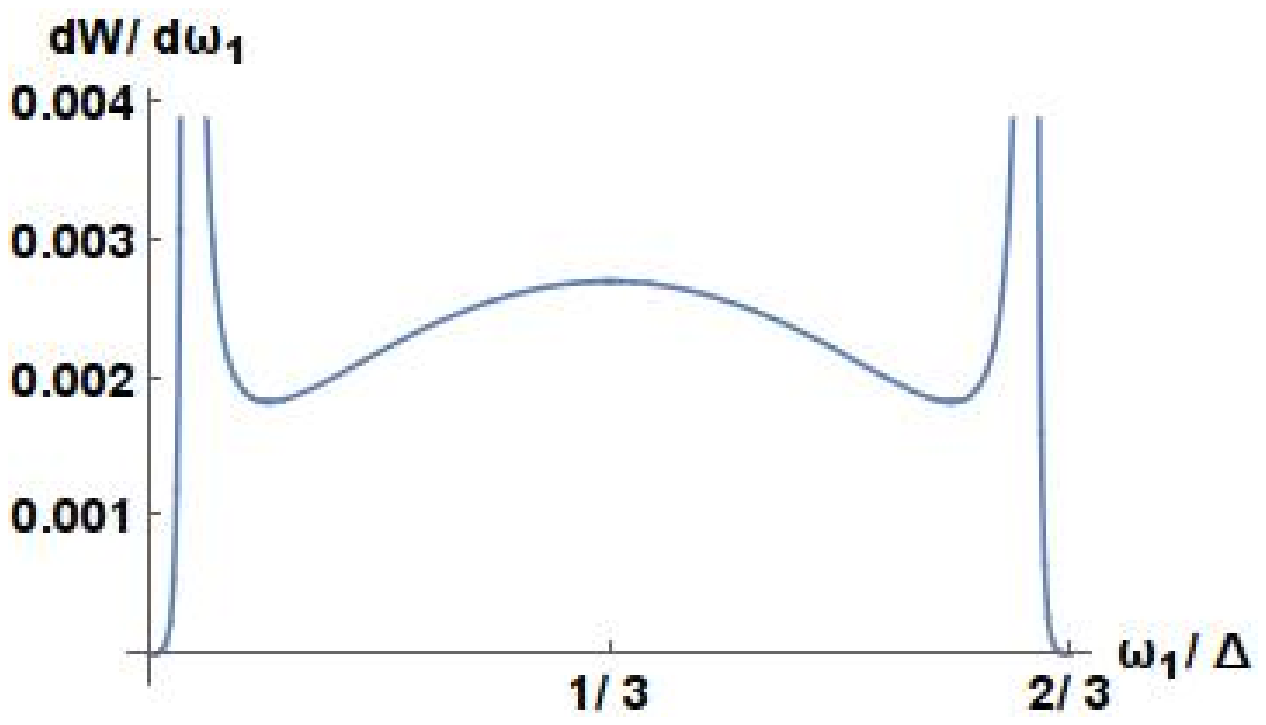}
\end{figure}

\end{document}